\RequirePackage{fix-cm}
\documentclass[twocolumn]{svjour3}          
\smartqed  
\usepackage{graphicx}
\usepackage{amsmath}
\usepackage{gensymb}
\usepackage{overpic}

\journalname{}
\def\makeheadbox{{%
\hbox to0pt{\vbox{\baselineskip=10dd\hrule\hbox
to\hsize{\vrule\kern3pt\vbox{\kern3pt
\hbox{\bfseries\ }
\kern3pt}\hfil\kern3pt\vrule}\hrule}%
\hss}}}

\def\makeheadbox{}

\begin{document}

\title{Heterogeneity Preserving Upscaling for Heat Transport in Fractured Geothermal Reservoirs
}
\subtitle{}


\author{Anna Nissen   \and   Eirik Keilegavlen \and Tor Harald Sandve \and Inga Berre \and Jan Martin Nordbotten}


\institute{A. Nissen \at
Universitety of Bergen \\
Department of Mathematics \\
P.O. Box 7803 \\
N-5020 Bergen, Norway \\
              Tel.: +047-47-55582857\\
              \email{anna.nissen@math.uib.no}           
}

\date{}

\maketitle

\begin{abstract}
In simulation of fluid injection in fractured geothermal reservoirs,
the characteristics of the physical processes are severely affected by the local occurence of connected fractures. To resolve these structurally dominated processes, there is a need to develop discretization strategies that also limit computational effort. In this paper we present an upscaling methodology for geothermal heat transport with fractures represented explicitly in the computational grid. The heat transport is modeled by an advection-conduction equation for the temperature, and solved on a highly irregular coarse grid that preserves the fracture heterogeneity. The upscaling is based on different strategies for the advective term and the conductive term, respectively. The coarse scale advective term is constructed from sums of fine scale fluxes, whereas the coarse scale conductive term is constructed based on numerically computed basis functions. The method naturally incorporates a coupling between the matrix and the fractures via the discretization, so that explicit transfer terms that couple solution variables in the fractures and the matrix are avoided. Numerical results show that the upscaling methodology performs well, in particular for large upscaling ratios, and that it is applicable also to highly complex fracture networks. 
\end{abstract}


\section{Introduction}
\label{intro}
Geothermal reservoirs are typically situated in igneous rocks, where the permeability of the reservoir mainly is governed by discrete fractures. The fractures occur on a range of scales, where large-scale connected structures tend to completely dictate flow paths. Flow and transport processes are strongly dominated by these structural heterogeneities. While large-scale fractures dictate preferential fluid pathways, fine-scale fractures contribute significantly to the subsurface heat exchange from the rock to the brine. The size of the fractures in a fractured reservoir typically follow a power law distribution, which means that at any scale of investigation, flow will be controlled by single fractures that cannot be represented by an upscaled permeability  \cite{BonnetEtAl2001}. Due to the lack of scale separation in fractured formations, particular care must be taken to appropriately account for local flow and transport characteristics.

Modeling approaches for fractured reservoirs can be classified into three main
categories based on their spatial representation in the considered
formation~\cite{DietrichEtAl2005}: single continuum models, multi-continuum
models, and discrete fracture-matrix models. Single continuum models average spatial
properties of the rock over grid cells that are much larger than the narrow
width of the structural heterogeneities (fractures), assuming that the concept
of representative elementary volume is valid. An advantage of this approach is
that standard reservoir simulators can be applied. In multi-continuum models,
the different characteristics of flow in different structural components are
preserved for each region of rock; see, e.g., \cite{Berkowitz2002} and
references therein. Each structural component, e.g. fractures on different
scales and matrix, is modeled by a representative continuum, and interacts with
the other continua comprising the same region. In this way integral transport
behavior can be better captured compared to when using a single continuum model.
In contrast, discrete fracture network (DFN) models represent fractures
explicitly and solely considers flow in the network of explicily represented
fractures; see, e.g.,
\cite{WatanabeTakahashi1995,BruelCacas1992,HayashiEtAl1999}. Discrete
fracture-matrix (DFM) models conceptually represent a combination of DFN models
and multi-continuum models, with explicitly represented fractures between
regions of continuity (e.g.,
\cite{DietrichEtAl2005,ReichenbergerEtAl2006,MartinEtAl2005}), and is the
approach we consider here. The advantage of DFM models is that they allow
explicit representation of fractures that cannot be represented by an equivalent
continuum at the chosen scale in simulation models, while at the same time
account for (upscaled) permeability due to small scale fractures and pores in
the regions surrounding the fractures. Therefore, the DFM models account for the
lack of scale separation by modelling fractures that dominate flow
characteristics at a certain scale explicitly, while at the same time accounting
for the region outside these fractures being permeable. This is of particular
importance in geothermal heat transport, where the permeability of the region
surrounding the fractures with dominating flow plays an important role for
heat transport.

Discretization methods for DFM models are challenging due to the large difference in scales between fractured and non-fractured regions. Co-dimension one approaches treat fractures as lower dimensional objects in the geometrical domain, and associate the removed dimension with a hydraulic aperture in the computations. This simplifies modelling at the cost of a small volumetric error, see e.g. \cite{Karimi-Fard2004,SandveEtAl}. The explicit representation of fractures in DFM models necessitate large flexibility in terms of grid generation. Frameworks that slightly modifies the fracture network to align with a background grid have been developed to avoid excessive grid refinement close to e.g. fracture intersections \cite{MusthaphaMusthapha2007,Karimi-FardDurlofsky2016}. In addition to gridding and discretization challenges, numerical representation using DFM models typically leads to a high number of degrees of freedom compared to using continuum models. Efficient numerical methods are therefore important.

In this paper we present a heterogeneity-preserving upscaling method for advective-conductive heat transport in fractured reservoirs. 
Our main contribution is two-fold:
First, we introduce an upgridding method that is tailored to drainage of 
energy from the matrix to the fracture system.
Second, we study discretization schemes for heat transport on the resulting coarse grid.
Following \cite{Hauge2012}, the coarse grid used for heat transport is
based on a fine-scale velocity field, which is assumed to be known.
The coarse cells are constructed by merging fine-scale cells with similar
flow properties and distance from the fracture network. 
The resulting coarse grid is well suited to represent both the slow drainage of heat
from matrix to the fracture network, and the fast transport within the network.
We emphasize that the user interacts with the coarsening algorithm only via 
a few meta-parameters.

Depending on the geometry of the fracture network, the coarse grid can have 
cells with highly irregular shapes, including non-convexity and high aspect
ratios. 
While the advective term can be discretized from the known fine-scale velocity
field in the manner of \cite{Hauge2012}, the complex grid poses challenges for
the discretization of the heat conduction term.
To overcome this, we use ideas from multiscale methods developed for 
elliptic and parabolic equations.
Multiscale methods have been used extensively
in porous media applications to efficiently discretize the pressure equation in
a way that accurately represents fine-scale heterogeneities, see, e.g.,
\cite{Jenny2003,ZhouTchelepi2008,Hajibeygi2011}.
Due to the emphasis on material heterogeneities, the majority of the work on multiscale methods applied to porous media problems has with a few exceptions been focused on structured grids.
The framework presented in \cite{Moyner2016}, on the other hand,
makes few assumptions with regards to grid geometry, and uses an
algebraic procedure to compute multiscale basis functions. 
This makes the framework in \cite{Moyner2016} well suited for general coarse grids, and we therefore use it as a basis for discretization of the coarse-scale conductive term.
However, the highly irregular shape of the coarse cells makes it necessary to
introduce modifications to avoid numerical instabilities.
We also study a non-conforming discretization of the coarse advection term, i.e. 
based on piecewise constant basis functions.
This second approach can be expected to have inferior approximation properties,
but it does provide a simple and robust alternative to the multiscale-based
discretizations.

To study the performance of our methods we consider fracture network ranging
from a Cartesian structure to a highly complex network that consists of a large number
of stochastic fractures.
We then study the accuracy of the methods under varying upscaling ratios and fluid injection rates.
The numerical examples show that both discretizations capture the main features of
the transport, compared to a fine-scale simulation. As expected, the
multiscale based method gives the most accurate predictions, in particular in
regions where conduction dominates.

The outline of the paper is as follows. In section \ref{equations} we describe the governing equations for flow and transport in geothermal reservoirs and the discretization on the fine scale. The framework for coarse scale discretization is presented in section \ref{upscaling}. Numerical experiments are presented in section \ref{num_exp} and the paper ends with conclusions and a discussion in section \ref{concl}.

\section{Fine Scale Model}
\label{equations}
In this section we present the governing equations for flow and heat transport. We also derive the fine scale discretization that later will be the starting point for our upscaling methodology.

\subsection{Flow and Heat Transport Equations}
\label{sec:goveq}
We consider single phase incompressible flow. Mass conservation of the fluid, 
\begin{align}
\nabla \cdot \textbf{v} = q, 
\label{mass_conservation}
\end{align}
in combination with Darcy's law,
\begin{align}
\textbf{v}  = - \frac{\textbf{K}}{\mu} \nabla p,
\label{Darcys_law}
\end{align}
is combined into an elliptic equation for the pressure, $p$,
\begin{align}
- \nabla \cdot \left( \frac{\textbf{K}}{\mu} \nabla p \right) = q,
\label{pressure_eq}
\end{align}
where $\textbf{K}$ is the permeability tensor, $\mu$ is the viscosity, which 
is assumed to be constant, 
$\textbf{v} $ is the Darcy flux field and $q$ contains any sources or sinks. 
The relation between hydraulic aperture, $a$, and fracture permeability, $k_f$, is 
modelled as 
\begin{align}
   k_f = \frac{a^2}{12}.
\label{frac_perm}
\end{align}


The effective volumetric heat capacity for rock and fluid combined is given by 
\begin{align}
(\rho c_p)_{eff} = \phi (\rho c_p)_f + (1 - \phi) (\rho c_p)_r,
\end{align}
where subscripts $f$ and $r$ denote fluid and rock, respectively, and $eff$ denotes effective. Here, $\rho$ is the density, $c_p$ is the specific heat capacity, and $\phi$ is the porosity. We assume that the specific heat capacities of fluid and rock varies slowly with time. Assuming local thermal equilibrium for a single phase flow, the energy equation that describes heat tranfer through the reservoir can be simplified to a linear advection-conduction equation for 
the temperature, $T$, given by 
\begin{align}
(\rho c_p)_{eff} \frac{\partial T}{\partial t}  +  (\rho c_p)_f  \nabla \cdot (\textbf{v}  T) - \nabla \cdot ({C} \nabla T) = q_e.
\label{temperature_eq}
\end{align}
In addition, $q_e$ is the energy source term, and $C$ is the effective thermal conductivity for the porous medium saturated with fluid at local thermal equilibrium.
For simplicity, we will assume a constant value for conductivity. The assumptions made above are reasonable for water-filled geothermal reservoirs, including a wide range 
 of temperatures for enhanced geothermal systems (EGS) situated at large 
 depths. In deep reservoirs water remains a compressed liquid for high 
 temperatures due to high hydrostatic pressures \cite{Cheng79}. We obtain a decoupled system of equations \eqref{pressure_eq} and 
 \eqref{temperature_eq} that together with appropriate boundary conditions can 
 be solved sequentially in numerical simulations. The coupling is one-way via the Darcy flux, $\textbf{v}$, due to the assumptions of an incompressible fluid with constant viscosity, and the 
pressure equation is only solved once, at the start of the simulation. 


The ratio between heat advection and conduction is commonly characterized by 
the heat P\'eclet number $Pe$, defined as 
\begin{align}
Pe = \frac{L \cdot u  \cdot (\rho c_p)_{eff}}{\alpha},  
\label{Pe_number}
\end{align}
where $L$ is a representative length scale, typically the distance between 
injector and producer wells, $u$ is an average velocity, and $\alpha$ is the 
average heat conductivity.  
The flow injection rate generally determines the flow rate through the 
fracture network and subsequently the highest value of $Pe$ in the reservoir. Depending on e.g. the matrix-fracture permeability contrast, the heat advection in the matrix can be considerably slower compared to that in the fractures leading to a second response from advection through the matrix \cite{Pruess1983}. As an effect the values of $Pe$ can, depending on the local flow field, vary by orders of magnitude in different regions of the reservoir. 

\subsection{Fine Scale Discretization}
\label{fine_discr}
We use an unstructured computational grid, constructed as a triangulation constrained to the 
fractures, with a slightly modified version of the algorithm presented in 
\cite{Holm2006}. 
Equations \eqref{pressure_eq} and \eqref{temperature_eq} are discretized using 
a finite volume discretization. Let \boldmath{$\omega$} be a set of fine grid 
cells, \boldmath{$\omega$} = \unboldmath\{$\omega_i\}, i=1,\cdots, N_f$, where $N_f$ is the total number of fine cells. 
Integrating equation \eqref{pressure_eq} over each grid cell $\omega_i$ and 
using the divergence theorem gives 
\begin{align}
\int_{\partial \omega_{i}} \textbf{v} \cdot \textbf{n} \hspace{3pt} dS =  
\int_{\omega_i} q \hspace{3pt} dV,  \hspace{10pt}   i=1,\cdots N_f, 
\label{pressure_eq_2}
\end{align}
where $\partial \omega_{i}$ is the boundary of grid cell $\omega_i$ and 
$\textbf{n}$ is the outward normal of $\partial \omega_{i}$. 
Similarly, equation \eqref{temperature_eq} becomes
\begin{align}
(\rho c_p)_{eff} \int_{\omega_i} \frac{\partial T}{\partial t}  \hspace{3pt} dV  -  (\rho c_p)_f  \int_{\partial \omega_{i}} (\textbf{v}  T) \cdot \textbf{n}  \hspace{3pt} dS \notag \\
+ \int_{\partial \omega_{i}} ({C} \nabla T) \cdot \textbf{n}  \hspace{3pt} dS =  \int_{\omega_i} q_e \hspace{3pt} dV, \hspace{10pt}   i=1,\cdots N_f.
\label{temperature_eq_2}
\end{align}
A two-point flux approximation is used to discretize both equation 
\eqref{pressure_eq_2} and the conductive part of equation 
\eqref{temperature_eq_2}. The fractures are incorporated by the hybrid 
discretization developed in \cite{Karimi-Fard2004}, where the permeability 
heterogeneity is preserved by the introduction of cells along the fractures. 
The advective fluxes in equation \eqref{temperature_eq_2} 
are approximated using standard upstream weighting \cite{AzizSettari1970}.

The discretization of equation \eqref{pressure_eq} gives rise to a linear system on the form
\begin{align}
\textbf{A}_{f} \textbf{p}_f = \textbf{q},
\label{discrete_pressure}
\end{align} 
where $\textbf{p}_f$ is a vector containing cell-centered pressure values. 
Equation \eqref{discrete_pressure} can be solved either via a direct solver or by using an appropriate iterative solver, see, e.g.,  \cite{SandveWRR}. 
Similarly, after discretizing equation \eqref{temperature_eq_2} in space, the semi-discrete temperature equation 
reads  
\begin{align}
 \frac{d \textbf{T}_f}{d t} + (\textbf{A}_{f,conv} + \textbf{A}_{f,cond}) \textbf{T}_f = \textbf{q}_e,
\label{semi_discr_temp}
\end{align}
where $\textbf{T}_f$ is a vector containing cell-centered temperature values, $\textbf{A}_{f,conv}$ and $\textbf{A}_{f,cond}$ approximate the advective 
and the conductive terms in eq. \eqref{temperature_eq_2}, respectively, and $\textbf{q}_e$ approximate the right hand side. $\textbf{A}_{f,conv}$, $\textbf{A}_{f,cond}$, and $\textbf{q}_e$ are all scaled 
with the inverse of the effective heat capacity. 


For time propagation we use an implicit second order accurate backward differentiation formula (BDF2), which has favorable stability properties, with time steps sufficiently small for temporal diffusion not to 
pollute the simulation accuracy. 



\section{Heat Transport Upscaling}
\label{upscaling}


The fine scale discretization with explicitly represented fractures in the
computational grid is often too computationally expensive to use directly in the
heat transport simulations.  As an alternative approach, the fine scale
discretization can be used as the basis for numerical upscaling, where most of
the computations are carried out on a coarser grid. In this section, we
introduce an upscaling method for simulation of heat transport in fractured
media. The upscaled transport model requires the construction of a coarse grid,
together with a discretization of equation \eqref{temperature_eq_2} on that
grid. We assume that a fine-scale velocity field is available, so that fine
scale heterogeneities in the flow model can be incorporated into the coarse
scale transport discretization. If the coupling between flow and transport is
assumed to be weak, a single fine scale pressure solve may be sufficient, in
which case the pressure solve will not constitute a large part of the overall
simulation cost.  Alternatively, iterative solvers or multiscale methods
tailored for fractured media can be applied to effectively produce conservative
approximations to the velocity field on the fine scale, see for example
\cite{SandveEtAl,Shah2016,Tene2015}. Because of the assumption of incompressible
flow, we use a direct fine scale pressure solver in this work. Since no
additional errors from an approximate velocity field are introduced, errors due
to the coarse scale transport solver can be isolated, and the performance of the
method can be thoroughly investigated.


\subsection{Coarse Scale Grid Construction}
\label{upscaling_grid}
To design adequate coarse grids, insight into the behavior of transport
processes in fractured media is needed.  Global transport is often dominated by
flow in the fractures, while heat stored in the matrix is transported into the
fracture network by a combination of matrix flow and conduction. This suggests
two criteria for the construction of coarse grids. First, the coarse grid should honor the structure of the flux field to adapt to large scale connectivity, i.e. large scale fractures. Specifically, we have found it useful
to apply the flow-based indicator framework introduced by Aarnes et al.
\cite{AarnesHaugeEfendiev2007}, and to use time-of-flight (TOF) as indicator
function for the coarse grid. Second, the assumption that heat drains from the
matrix to nearby fractures suggests that the coarse cell geometry in the matrix
should reflect the distance to nearby fractures. We achieve this by classifying
fine scale cells located in the matrix according to their geometric distance
from the closest fracture and refining the coarse grid according to the distance
measure. It is instructive to draw the parallel between this second criterion
and the so-called multiple interactive continuum (MINC) models
\cite{Pruess1985}.
Both models use distance from fractures as a proxy for the time it takes for
heat to drain towards the fractures.
A major difference is that since our starting point is a fine-scale grid,
estimation of interaction between coarse cells is relatively simple even for 
general fracture geometries.
Moreover, our approach can readily be combined with heat advection.

\begin{figure*}
\centering
\includegraphics[width=7cm,height=7cm]{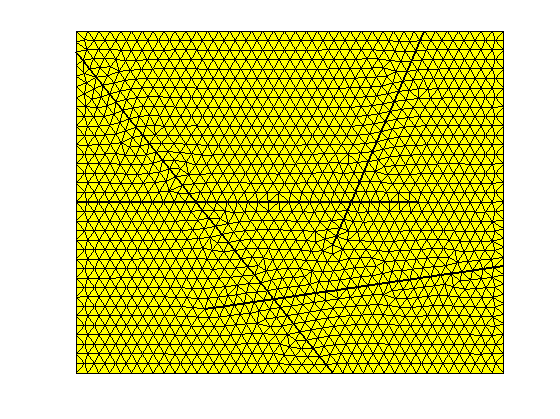}
\includegraphics[width=7cm,height=7cm]{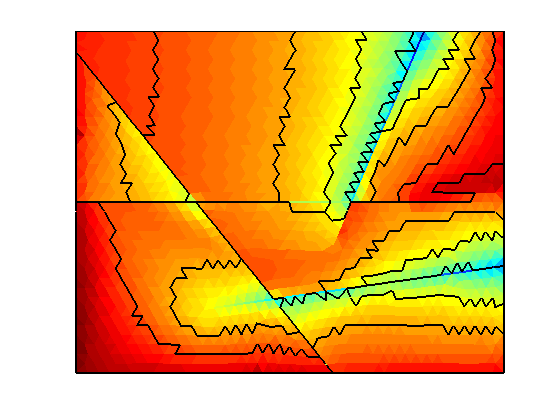}\\
\includegraphics[width=7cm,height=7cm]{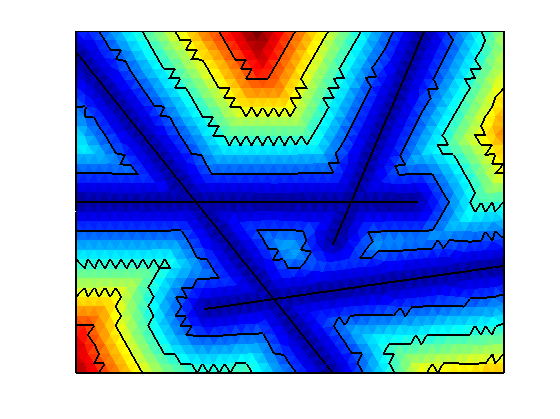}
\includegraphics[width=7cm,height=7cm]{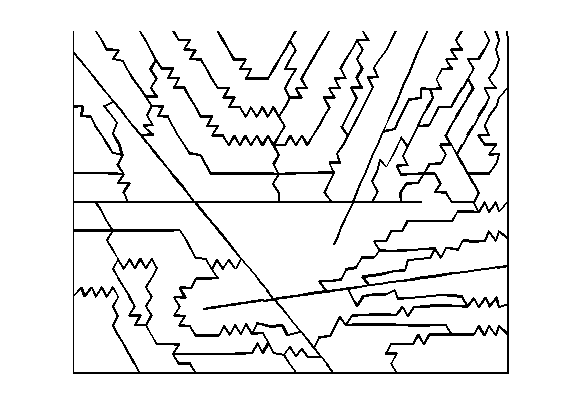}
\caption{Construction of the coarse grid. Fine scale grid with fractures explicitly represented (top left). TOF indicator and outlined coarse grid based on TOF indicator (top right). Distance indicator and outlined coarse grid based on distance indicator (bottom left). Final coarse grid, where the TOF coarse grid and the distance coarse grid have been combined (bottom right).}
\label{coarse_grid_construction}
\end{figure*}

We refer to the
mapping from the fine grid to the coarse grid as a \emph{coarse grid partition},
i.e. the description of which fine grid cells constitute each coarse grid cell.
The different steps in the construction of the coarse grid are visualized in
figure \ref{coarse_grid_construction} and briefly described in the following
steps:

\begin{enumerate}
\item Generate an initial coarse grid partition of a  fine grid based on a coarsening indicator, e.g., based on TOF.
\item Refine the initial coarse grid using a distance based grid partition. This partition is defined by a predefined coarse grid resolution parameter and a distance indicator (measuring the distance between the cell center of a fine cell and the closest fracture). 
\item Hybrid coarse cells (consisting of both fine fracture cells and fine matrix cells) are separated so that all coarse cells either contain only fine fracture cells or fine matrix cells.
\item Coarse cells are ensured to be (or split into) connected components.
\end{enumerate} 

The coarse cells can be merged and refined in an iterative loop in order to
obtain a coarse grid with preferred properties \cite{Hauge2012}. The focus of this work is not to
obtain an optimal coarse grid, but rather to design a coarse scale
discretization that works well also for highly irregular coarse grids. We remark
that the coarse grid in figure \ref{coarse_grid_construction} gives a fair
representation of the grid cell shapes we have encountered. For transport
problems dominated by fracture flow, the number of grid cells can be vastly
reduced and still capture large scale behavior for a coarse grid constructed
based on flow indicators compared to when a more structured coarse grid is used.
However, with a more irregular coarse grid the discretization on the coarse
scale needs to be carefully constructed. This is described in the following
sections.

\subsection{Coarse Scale Discretization}
\label{upscaling_discr}
 Let \boldmath{$\Omega$} be a set of coarse grid cells, \boldmath{$\Omega$} = \unboldmath\{$\Omega_{\ell}\}$, $\ell=1,\cdots, N_c$, where $N_c$ is the total number of coarse cells. The coarse scale finite volume discretization takes the same form as the fine scale discretization in equation
\eqref{temperature_eq_2}, with integration over a coarse cell $\Omega_{\ell}$. 

In the following, we detail the coarse scale discretization of the advective
and conductive terms, respectively.
As we will see, the former can be treated relatively easily, while the latter requires
more attention.

\subsubsection{Coarse Scale Advective Term}
The treatment of the coarse scale advective term assumes that fine scale fluxes are known. Following \cite{Hauge2012}, a control volume discretization of the advective term is obtained by integrating the (known)
fine scale fluxes to net fluxes on the coarse scale,
\begin{align}
 \int_{\partial \Omega_{\ell}} (\textbf{v}  T) \cdot \textbf{n}  \hspace{3pt} dS \approx \notag \\
\sum_{k \neq \ell}  \max \left( T_{\ell} \sum_{\gamma_{ij} \in \Gamma_{k\ell}} 
\textbf{v}_{ij} -  T_k \sum_{\gamma_{ij} \in \Gamma_{k\ell}}  \textbf{v}_{ij} 
\right),
\label{net_fluxes}
\end{align}
where $\gamma_{ij}$ denotes a face between fine grid cells $i$ and $j$, and $\Gamma_{k\ell}$ a face between coarse grid cells $k$ and $\ell$.




\subsubsection{Coarse Scale Conductive Term}
For the coarse scale conductive term we take advantage of the fact that the same structure is found in elliptic and parabolic pressure equations. In porous media applications, coarse scale discretizations of elliptic equations with strongly variable coefficients (related to heterogeneous and anisotropic media) has received much attention the last decade in the context of multiscale methods \cite{Jenny2003,AarnesEfendiev2006,ZhouTchelepi2008,Hajibeygi2008}. Due to the strong hyperbolic component of the heat transport, a projection of the full temperature profile from the coarse to the fine scale is highly challenging, and we therefore cannot directly apply a multiscale framework to the advection-conduction equation.  Nevertheless, we construct the upscaled discretization of the conductive term inspired by multiscale metods.  


The coarse scale finite volume discretization matrix for the conductive term, 
$\textbf{A}_{c,cond}$, is related to the corresponding fine scale discretization matrix, $\textbf{A}_{f,cond}$, by
\begin{align}
\textbf{A}_{c,cond} = \textbf{R} \textbf{A}_{f,cond} \textbf{P},
\label{coarse_scale_diff_term}
\end{align} 
where $\textbf{R}$ is a restriction operator represented by a matrix of size $N_c\times N_f$, and $\textbf{P}$ is a prolongation matrix of size $N_f\times N_c$. The restriction operator, $\textbf{R}$, maps the fine scale discretization to the coarse scale. We use a finite volume restriction matrix, with elements given by 
\begin{align}
\textbf{R}_{i,j} = \left\{ \begin{tabular}{cc}
1 & if  $\omega_j \in \Omega_i$,  \\
0 & otherwise,
\end{tabular} \right.
\end{align}
see \cite{ZhouTchelepi2008}.
$\textbf{P}$ can be expressed as a column matrix,
\begin{align}
\textbf{P} = \left[ P_1 \quad P_2 \quad \cdots \quad P_{N_c} \right],
\end{align} 
where the columns $P_i$, $i=1,\cdots, N_c$, contain prolongation operators or \emph{basis functions}. 

A coarse scale discretization on the form \eqref{coarse_scale_diff_term} is often used for multiscale methods for elliptic equations with variable coefficients, see for instance \cite{ZhouTchelepi2008,Moyner2016}. In that case, the basis functions can be combined to directly map the coarse scale solution variables to a fine scale approximation. For elliptic equations that stems from strongly heterogeneous and anisotropic porous media problems, a main difficulty with respect to multiscale methods is to define basis functions that give a numerical solution with desired properties. More generally, the construction of the prolongation matrix $\textbf{P}$ is a long-standing issue in the field of coarse discretizations, and it is also closely related to the formulation of coarse spaces in the domain decomposition and multigrid literature \cite{Toselli2005}.

It is a common feature of many multiscale methods and numerical upscaling methods that they are most easily realized on relatively structured grids. As exemplified in figure \ref{coarse_grid_construction}, the grid coarsening strategy presented in Section \ref{upscaling_grid} can give rise to highly irregular coarse grids. To define a discretization of the conductive term on these grids based on geometric arguments is very challenging. Instead we consider an algebraic construction of the basis functions in the coarse discretization presented in \cite{Moyner2016}, which makes the method much more amenable to complex coarse grids. The algebraic construction is closely related to ideas in smoothed aggregation algebraic multigrid methods \cite{Vanek1996,Nordbotten2008}. In the following subsections, we describe how the methodology in \cite{Moyner2016} is applied and modified to obtain a suitable discretization of the conductive term in our advection-conduction equation.

\subsubsection{Interaction Regions}
\begin{figure*}
\centering
\includegraphics[width=5.6cm,height=5.5cm]{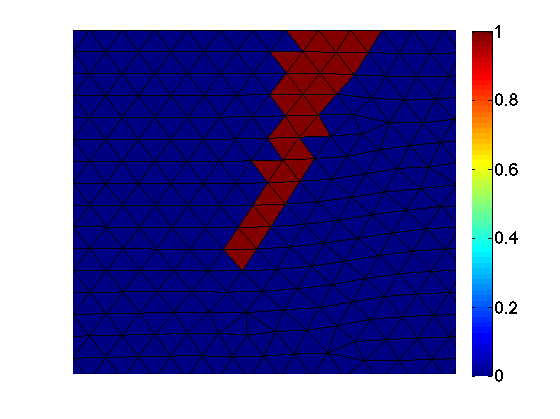}
\includegraphics[width=5.6cm,height=5.5cm]{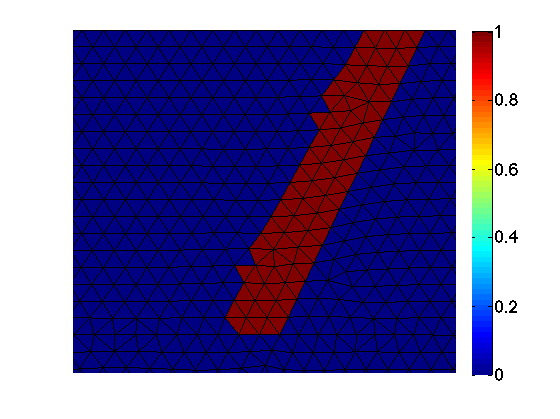}
\includegraphics[width=5.6cm,height=5.5cm]{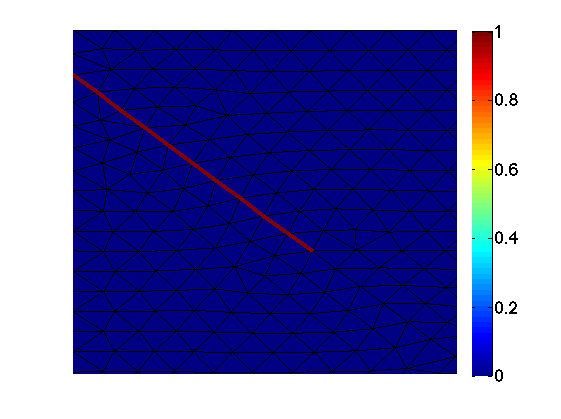} \\
\includegraphics[width=5.6cm,height=5.5cm]{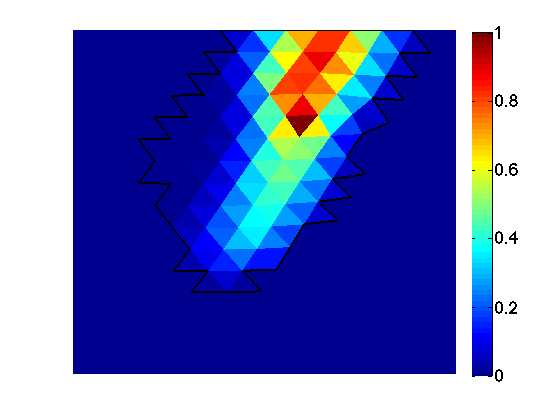} \hspace{-2pt}
\hspace{-2pt} \includegraphics[width=5.6cm,height=5.5cm]{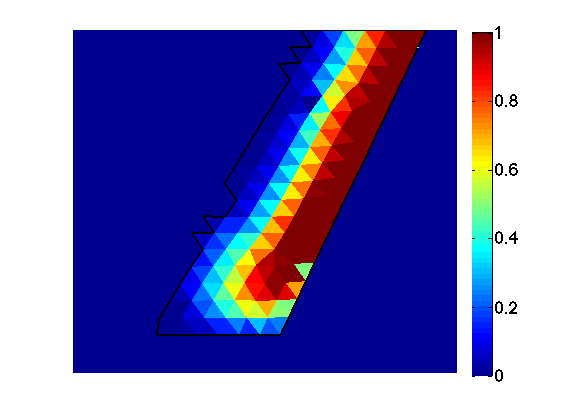} \hspace{-3pt}
\includegraphics[width=5.6cm,height=5.5cm]{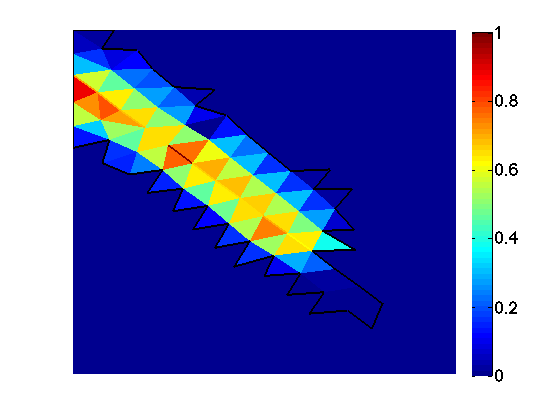}
\caption{Initial (constant) basis functions for three coarse cells from the coarse grid in figure \ref{coarse_grid_construction} are shown in the top figures. The corresponding smoothed basis functions after termination of iterations are shown in the bottom figures after 9, 50, and 9 iterations, respectively. The boundaries of the interaction regions for the basis functions are depicted with black lines in the bottom figures.}
\label{basis_functions}
\end{figure*}

The region of support for a basis function is defined in terms of interaction regions.
These regions resemble discretiztion stencils for multipoint flux approximations \cite{Aavatsmark2002} in the sense that all coarse cells that share a vertex with coarse cell $i$ contain some fine cells that are included in $P_i$.
Conceptually, we need to consider three different types of interactions: 
1) matrix cells neighboring matrix cells,
2) matrix cells neighboring fracture cells, and 
3) fracture cells neighboring fracture or matrix cells. 
Interactions regions for these three types are illustrated in figure \ref{basis_functions}, together with the corresponding coarse cells and basis functions for these cells, respectively.
The interaction regions are constructed by the algorithm in \cite{Moyner2016}, with some adaptions to account for the presence of fracture cells.
Specifically, we do not allow for interaction between coarse matrix cells that are separated by a fracture cell. This is due to the assumption of advective-dominated heat transport in the fractures.

\subsubsection{Algebraic Smoothing}
\label{sec:smoothedbasis}
Having defined the support of the basis functions in terms of the interaction regions, we go on to explicitly compute the basis functions. We use the algebraic construction from \cite{Moyner2016}, described briefly below. The initial description of the elements in the prolongation matrix $\textbf{P}$ is
\begin{align}
\textbf{P}_{i,j}^{(0)} = \left\{ \begin{tabular}{cc}
1 & if  $\omega_i \in \Omega_j$, \\
0 & otherwise,
\end{tabular} \right.
\end{align}
i.e. $\textbf{P}^{(0)} = \textbf{R}^T$.
We note that $\textbf{P}^{(0)}$ is a partition of unity, and forms a $P_0$ (piecewise constant) discretization on the coarse grid.

Now, successive iterations for basis function $P_i = \textbf{P}_{\cdot,i}$ are given by
\begin{align}
P_i^{(n+1)} = P_i^{(n)} - \omega \textbf{D}^{-1} \textbf{A}_{f,cond} P_i^{(n)},
\end{align}
where $\textbf{D} = diag( \textbf{A}_{f,cond})$, $\omega \in (0, 1]$ is a relaxation parameter, and $n$ denotes the iteration number. 
The effect of the iterations is to smoothen the basis functions.
The support of a basis function is limited to its interaction region, this is enforced after each iteration by 
simply truncating values that fall outside the interaction region. 
However, the truncated basis functions will generally not form a partition of 
unity. As a remedy, the functions are rescaled cell-wise to ensure 
that they sum to one within each fine scale cell \cite{Moyner2016}.

\subsubsection{Non-Oscillatory Basis Functions}
\label{monotonicity}
The above construction of basis functions works well for coarse cells with a
reasonably regular geometry. 
However, we have observed that when the coarse grid becomes highly irregular,
as illustrated in figure \ref{coarse_grid_construction}, the basis functions
may have a non-monotonic behavior, which consequently lead to oscillations in
the numerical solution.
To see why the non-monotonic profiles occur, we note that for cells with large
aspect ratios or with complex shapes, interaction regions associated with neighboring cells are not necessarily in contact
with the cell center. 
Thus the rescaling to preserve a partition of unity, which is carried out
for each interaction region individually, can cause the basis function not
to be monotonically decreasing from the cell center. In this and the following subsection we discuss two approaches to mitigate this effect.

\begin{figure}
\includegraphics[width=7cm,height=7cm]{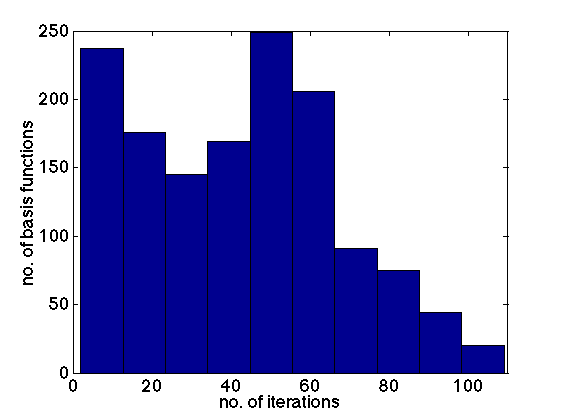}
\caption{Histogram over the iteration count for basis functions in the simulations in section \ref{Pe_exp}. There are 1412 of a total of 1852 basis functions for which iterations are terminated early. 440 basis functions are still updating when the smoothing is terminated based on the residual, and therefore not included in the histogram.}
\label{iteration_histogram}
\end{figure}

As a partial remedy to the oscillations we limit the number of iterations in the construction of basis
functions by minimizing an energy functional \cite{vanLent2009}. Specifically, we define $E_i$ as the discrete energy 
of the basis function for coarse grid cell $i$, $\Psi_i$ by 
\begin{align}
E_i = P_i^T \textbf{A}_{f,cond} P_i.
\end{align}
Oscillations in the basis function will increase the discrete energy, and we 
therefore terminate the iterations for basis function $P_i$ if for some 
iteration $n$ we have 
\begin{align}
E_i^{(n+1)}  > E_i^{(n)},
\label{energy_criteria}
\end{align}
while continuing to update basis functions that have not been terminated. 
The termination freezes the basis functions in all fine-scale cells contained 
within the interaction region of coarse cell $i$.
While the termination criterion \eqref{energy_criteria} does not guarantee
monotone basis functions, the approach has been sufficient for the 
fracture networks considered here. In addition to the termination condition
\eqref{energy_criteria} for each basis function, global criteria for terminating
the iterations are based on the size of the residual and by a user given total
number of iterations. Figure \ref{iteration_histogram} shows the distribution of basis functions
terminated at different numbers of iterations for the fracture network considered in section \ref{Pe_exp}. When the smoothening is terminated based on the residual after $111$ iterations, 24$\%$
($440/1852$) of the basis functions are still updating, and they are therefore
not included in the figure.

\textbf{Remark} In addition to non-monotonic basis functions, excessive
smoothing may also result in eigenvalues of $\textbf{A}_{c,cond}$ with negative
real parts. This consequently leads to blow up of the numerical solution as time
propagates. If needed, this can be avoided by introducing an additional criteria
on the diagonal elements of $\textbf{A}_{c,cond}$ such that
$\textbf{A}_{c,cond,ii}>0$, $i =1, \cdots, N_c$. If the diagonal elements of
$\textbf{A}_{c,cond}$ are strictly positive, then by the Gershgorin circle
theorem the eigenvalues have real parts that are strictly positive.

\subsubsection{Piecewise Constant Basis Functions}
\label{sec:constbasis}
As noted above, oscillations in the basis functions can to a large degree be
avoided by reducing the number of smoothing iterations.
It is of particular interest to investigate the limit option of taking no
iterations at all, and thus represent the coarse-scale temperature with
piecewise constant basis functions.
This approach avoids the oscillations, and as the interaction regions are no
longer required, the method is simpler to implement.
Moreover, it leads to a symmetric
discretization of the coarse conduction term, assuming that the fine scale discretization matrix, $\textbf{A}_{f,cond}$, is symmetric, since in this case $\textbf{P} =
\textbf{P}^{(0)} = \textbf{R}^T$, and thus 
$\textbf{A}_{c,cond}= \textbf{R} \textbf{A}_{f,cond} \textbf{R}^T$.
However, the piecewise constant representation of temperature implies that 
temperature gradients can only be present on the boundary between the
coarse-scale cells, where they are represented by the fine-scale discretization 
$\textbf{A}_{f,cond}$.
That is, the coarse-scale conduction retains a dependency on the fine scale grid
size $h$, rather than on a representative coarse grid size $H$.

As the above considerations show, the constant basis function will therefore
lead to inferior approximations, in particular for large upscaling ratios. The
effect will be most pronounced in regions where conduction is the dominant
transport mechanism. 
Nevertheless, the approach is interesting, in that it is considerably simpler
and more numerically robust than the smoothed basis functions.

\subsubsection{Relation to Other Methods}
It is of interest to discuss differences with related methods for coarse discretizations in fractured media.
Recently, Tene et al. \cite{Tene2015} and Shah. et al \cite{Shah2016} developed multiscale methods for solving the pressure equation in fractured media. In both these works the fracture networks and the matrix are separated into two independent coarse grids, and exchange between fractures and matrix are modelled via a Peaceman-type coupling term that is proportial to the difference between the matrix and fracture pressures \cite{Hajibeygi2011}. 
While this approach can be advantageous for the pressure equation with high permeability contrast between fracture and matrix, it is less relevant for heat conduction that has no abrubt jumps in the parameter field.

%

\section{Numerical Experiments}
\label{num_exp}
In this section we consider three different test cases to illustrate the
applicability and performance of the upscaling framework. Numerical experiments
are run both with a smoothed basis and with constant basis functions,
respectively. First, we present a refinement study for a structured fracture
network on an underlying Cartesian fine scale grid to investigate the practical
importance of sufficiently smooth basis functions. We follow by investigating
the performance of the numerical method for two more complex fracture networks
using unstructured triangular fine scale grids.

In the numerical simulations we use standard constant values for granite for the volumetric heat capacity of the rock and the thermal conductivity, $(\rho c_p)_r = 2170$ kJ/(m$^3$K) and ${C} = 2.1$ J$/$(mKs), respectively. For the volumetric heat capacity of the fluid and viscosity, $(\rho c_p)_f = 4180$ kJ$/$(m$^3$K), and $\mu=1$ cP are used, respectively. The porosity in the matrix is set to $\phi=0.001$. The reservoir domain is $\Omega = [0, 1000] \times [0, 1000]$ m$^2$  for all simulations. Initially the temperature in the reservoir is set to 100\degree C, and then water with temperature 20\degree C is injected somewhere in the reservoir. The precise injection and production setups as well as the fracture networks are described  for each case in subsections \ref{conv_study}, \ref{Pe_exp}, and \ref{complex_network}, respcetively. 

The total internal energy per grid cell at time $t$ is given by
\begin{align}
e_i(t) = \left[\phi (\rho c_p)_f + (1-\phi) (\rho c_p)_r\right] V_i T_i(t) \notag \\
= (\rho c_p)_{eff} V_i T_i(t),
\end{align}
where $i$ refers to the grid cell index.
The relative $\ell_2$ error over space in the energy at time $t$, $\varepsilon(t)$, is then computed as
\begin{align}
\varepsilon(t)= \frac{\sqrt{\sum_{i=1}^{N_f} |\tilde e_{c,i}(t) - e_{ref,i}(t)|^2}}{\sqrt{\sum_{i=1}^{N_f} |e_{ref,i}(t)|^2}},
\end{align}
where $\tilde e_{c}$ is the energy vector for the upscaled solution projected on the fine grid, given by $\tilde e_{c} = \textbf{R}^T e_{c}$, and $e_{ref}$ is the energy vector for a reference solution computed on the fine grid. $N_f$ refers to the total number of fine grid cells.

\subsection{Accuracy with Respect to Coarsening Ratio}
\label{conv_study}
We conduct a refinement study using a fine underlying cartesian grid to better
understand the practical importance of the iterative framework. The fracture
network consists of six fractures that are aligned with the grid, three in each
direction. First, a coarse mesh is constructed based on a partitioning of the
fine grid. Because of the structured nature of this fracture network we base the
initial coarse grid partition solely on the distance to nearest fracture. Using
a flow based partition leads to similar results. The upscaling ratio, or \textit{coarsening factor} between
the coarse and the fine grid is defined as $CF = N_f/N_c$. The coarsening factor is consecutively
increased by refining the fine grid by a factor two in each spatial direction, while the same coarse grid is used. The coarse grid contains $601$ grid cells and the fine grids contain
$6889$, $26569$, $104329$, and $413449$ grid cells, respectively, including
one-dimensional fracture cells. Water with temperature 20\degree C is injected
at a rate of 0.1 $dm^2/s$ in the middle of the domain (at $(500,500)$), and extracted
where each fracture meets the exterior boundaries. The total simulation time is
$T=30$ years.

\begin{figure}
\centering
\includegraphics[width=7.8cm,height=7cm]{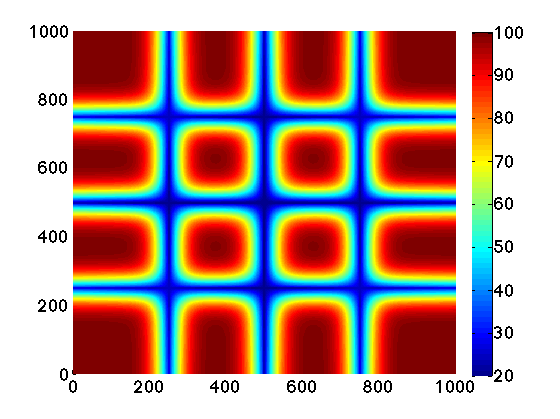}
\caption{Reference temperature profile from fine scale simulation in the refinement study presented in section \ref{conv_study}. The fine grid contains 413449 grid cells, including fracture cells.}
\label{ref_temp_conv_study}
\end{figure}

\begin{figure*}
\centering
\includegraphics[width=7.8cm,height=7cm]{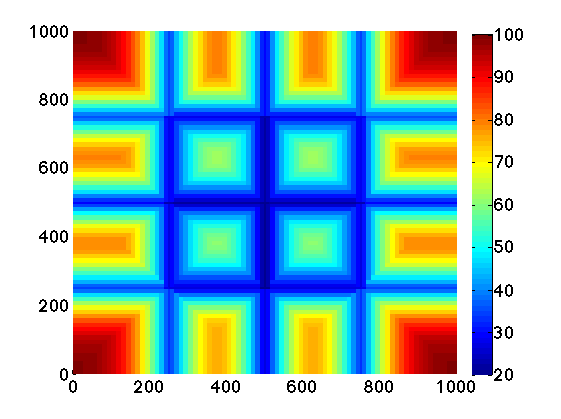}
\includegraphics[width=7.8cm,height=7cm]{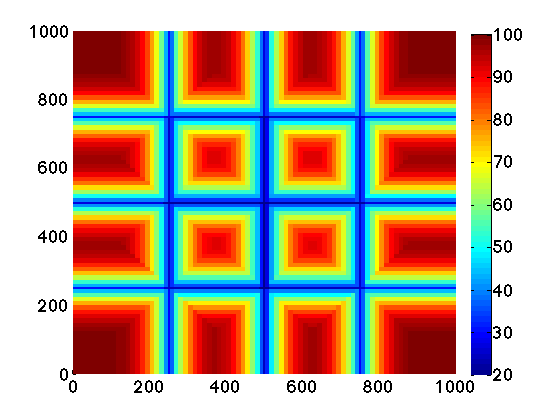}
\includegraphics[width=7.8cm,height=7cm]{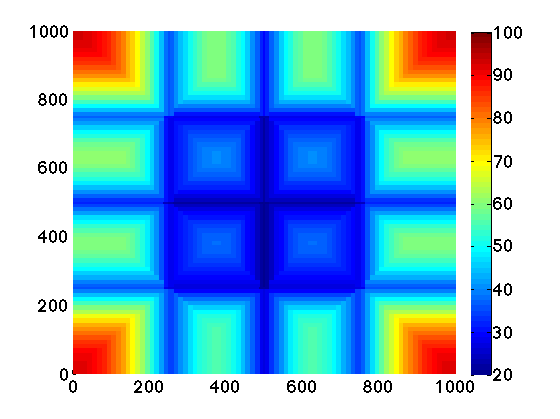}
\includegraphics[width=7.8cm,height=7cm]{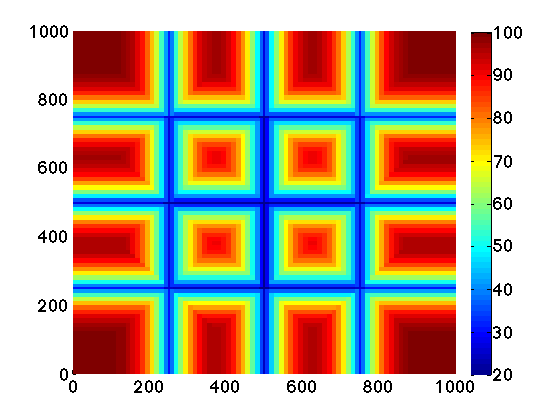}
\caption{Upscaled temperature profiles using piecewise constant basis functions (left) and smoothed basis functions (right), respectively, for the simulations in section \ref{conv_study}. The fine scale underlying grids have a total of $104329$ and $413449$ grid cells, respectively, depicted from top to bottom. The coarse grid contains $601$ grid cells.} 
\label{RunEx1_fine6889_MINC_MsRSB}
\end{figure*}

\begin{figure*}
\center
\includegraphics[width=7cm,height=7cm]{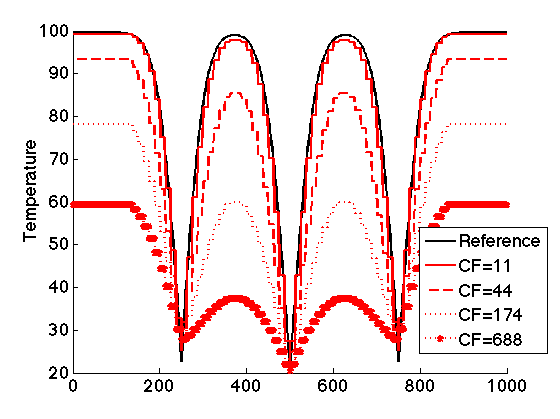}
\includegraphics[width=7cm,height=7cm]{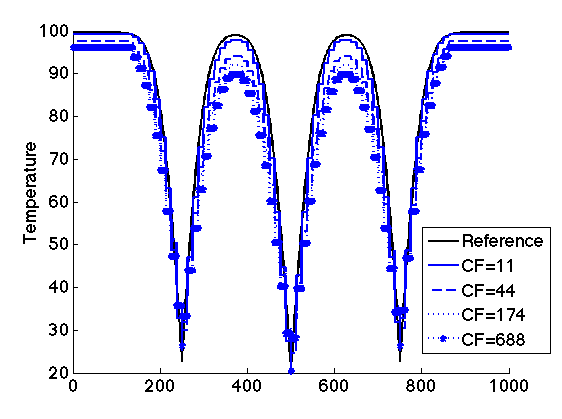}
\caption{One dimensional temperature profiles at $y=375$ for a successive grid refinement, with constant basis functions (left) and smoothed basis functions (right), respectively, for the refinement study in section \ref{conv_study}. The coarse grid contains $601$ grid cells for all simulations, the coarsening factors are for each case given in the figures.}
\label{1D_profiles_CartGrid}
\end{figure*}

The reference temperature for the finest grid is shown in figure
\ref{ref_temp_conv_study}. Upscaled temperature profiles for the two largest
refinement factors are shown in figure \ref{RunEx1_fine6889_MINC_MsRSB} for \\
smoothed basis functions and for constant basis functions, respectively. Figure
\ref{1D_profiles_CartGrid} shows one dimensional cross-sections of the
temperature profiles at vertical value $y=375$.
As seen from the figures, when using the smoothed basis functions, the upscaled method preserves the
main features of the fine-scale solution, even for high upscaling ratios.
The constant basis functions overestimate the diffusion, and 
the heat transfer from the matrix to the fractures is not well captured.
The error increases with the coarsening factor, as can also be seen in the energy
error given in table \ref{table_Ex1_cart}.
This is consistent with the discussion in section \ref{sec:constbasis}, in that
the gradient is approximated with respect to the fine-scale grid size $h$,
rather than the coarse resolution $H$.


To investigate the impact of smoothing iterations we have run a number of
simulations varying the maximal number of iterations, without terminating the
smoothing of any basis function early. The upscaled simulations in this section
are therefore run with a fixed number of iterations, given in table
\ref{table_Ex1_cart}. When the coarsening factor $CF$ is larger more iterations
are needed in order to reduce the error in the temperature equation. This is not
surprising since for each iteration the support of a basis function for a coarse
cell extends approximately one fine grid cell in each spatial direction.
Thus, for the temperature gradient to be taken with respect to $H$ rather than
$h$, the number of iterations should scale with $H/h$.
However, at a
certain number of iterations, depending on the refinement ratio, the error
typically starts to increase with more interations. This is due to that the
explicit representation of fractures leads both to a non-uniform coarse grid
structure and to irregular shapes of the interaction regions for each coarse
cell. What we have observed is that at a certain point more iterations introduce
oscillations in the the basis functions. To ensure the construction of (overall)
monotone basis functions, we enforce termination of the iterative procedure for
each basis function through the criteria of energy reduction described in
section \ref{monotonicity} in the numerical experiments presented in the
following sections.

\begin{table*}
\centering
\caption{Accuracy with respect to coarsening factor for the refinement study in section \ref{conv_study}.  $e_{CB}$ and $e_{SB}$ denote the $\ell_2$ energy error over space for constant and smoothed basis functions, respectively.}
\begin{tabular}{cccc}
\hline
No. fine cells & Coarsening factor $CF$ & $e_{CB}$ &$e_{SB}$ (no. of iterations) \\
\hline
6889 & 11 & $1.06 \cdot 10^{-2}$  & $1.06 \cdot 10^{-2} \hspace{5pt} (1)$   \\
26569 & 44 & $3.00 \cdot 10^{-2}$  & $1.63 \cdot 10^{-2} \hspace{5pt} (5)$   \\
104329 & 174 & $5.99 \cdot 10^{-2}$  & $1.90 \cdot 10^{-2} \hspace{5pt} (15)$   \\
413449 & 688 & $8.98 \cdot 10^{-2}$  & $2.15 \cdot 10^{-2} \hspace{5pt} (30)$   \\
\hline
\end{tabular}
\label{table_Ex1_cart}
\end{table*}


\subsection{Investigation of Heat Transport Characteristics}
\label{Pe_exp}
The previous example showed that the coarse discretization, with grids adapted
to the fracture network, combined with smoothed basis functions gives a good
approximation of the fine-scale temperature profiles.
When the conduction term is discretized with constant basis functions, heat
conduction is overestimated, and the approximation quality may suffer.
How severe this effect is naturally depends on the relative importance of heat
advection and conduction, as quantified by the heat P\'eclet number.

\begin{figure*}
\centering
\begin{overpic}[width=7cm,height=7cm]{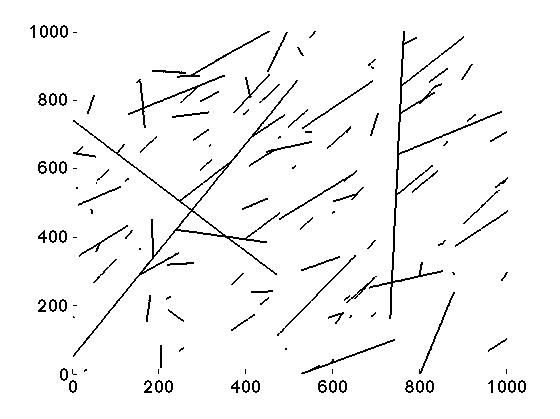} 
\linethickness{0.3mm}
\put(5,15){\vector(1,0){8}}
\put(-13,14){\text{Injection}}
\put(72,101){\vector(0,-1){8}}
\put(61,101){\text{Production}}
\end{overpic}
\includegraphics[width=7.8cm,height=7cm]{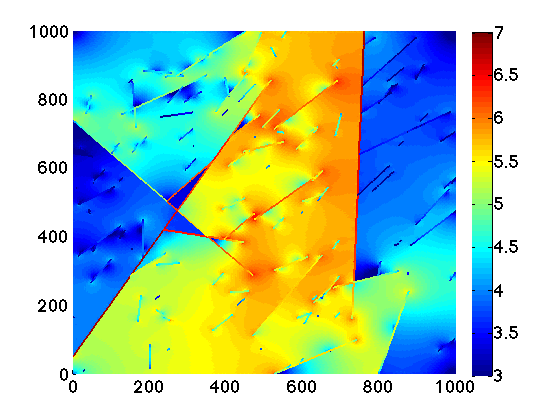} 
\caption{Fractures in the discretization for the simulations in section \ref{Pe_exp} (left) and the field with heat P\'eclet number per grid cell (right). The heat P\'eclet number plot is in logarithmic scale with base $10$. Injection and production locations are indicated by arrows in the fracture network plot.}
\label{Fractures_2}
\end{figure*}

In this set of numerical experiments we consider a more realistic setup and
investigate how variations of the heat P\'eclet number throughout the reservoir affect
the numerical results. We consider a cell based heat P\'eclet number, defined for each cell by equation \eqref{Pe_number}, where fine scale velocities averaged on each grid cell are used for $u$. The fracture network used in the simulations in this
section are generated using the stochastic fracture generator Frac3D
\cite{SilberhornHemminger2002}, except from a few of the largest fractures with
assumed known positions. The fracture network is shown in figure
\ref{Fractures_2}, and apertures range from $a=10^{-3}$ m to $a=0.1$ m for the
largest fractures. In practice, the largest of these values corresponds to fracture
corridors and fault zones rather than individual features, but we will for
simplicity refer to them as fractures. Note in particular that the large scale fractures do not form one single
connected network. The average heat P\'eclet numbers per fine grid cell are also shown in figure \ref{Fractures_2}. Note that the cell based heat P\'eclet numbers vary by orders of magnitude in different regions of the reservoir due to the fracture-matrix heterogeneity. From figure \ref{Fractures_2} we observe that, as expected, advection is the dominant process in the fractures.
Since the fractures do not form a continuous pathway between inlet and outlet,
the fluid needs to flow through the matrix.
This is reflected in relatively high flow rates in the matrix in the middle
of the domain, whereas the flow is smaller towards the domain boundary.
Based on the previous example, we expect the largest differences
between the smoothed and constant basis functions  in regions where the heat P\'eclet numbers attain low values.

\begin{figure}
\centering
\includegraphics[width=7cm,height=7cm]{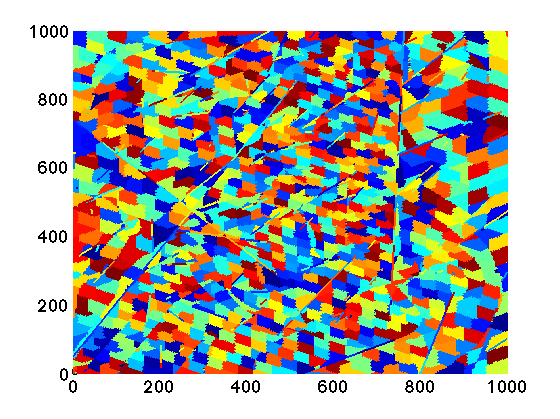}
\caption{The coarse grid partition used for the numerical experiments in section \ref{Pe_exp}. The coarse grid has 1852 grid cells and is generated from a fine grid with 103893 grid cells.}
\label{grid_partition}
\end{figure}

\begin{figure*}
\center
\includegraphics[width=5cm,height=5cm]{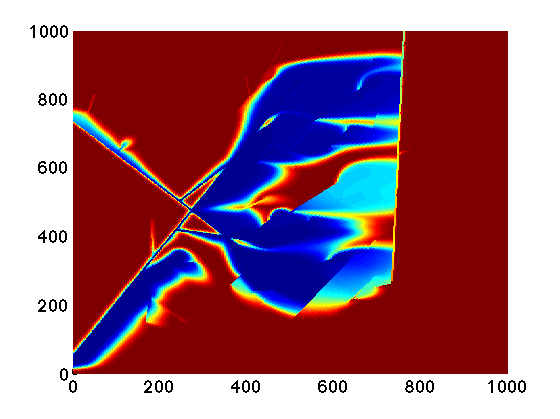}
\includegraphics[width=5cm,height=5cm]{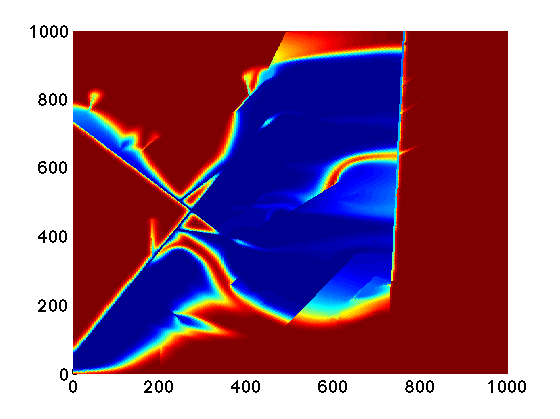}
\includegraphics[width=5cm,height=5cm]{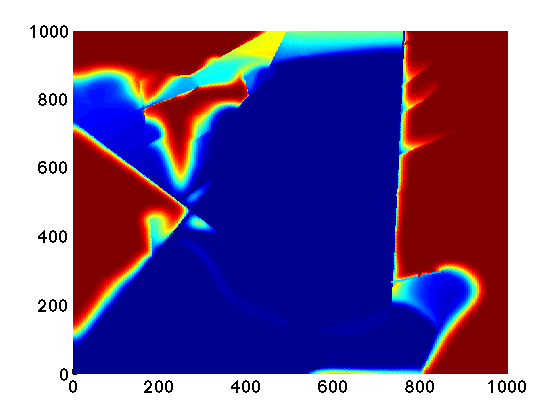}
\includegraphics[width=0.8cm,height=4.8cm]{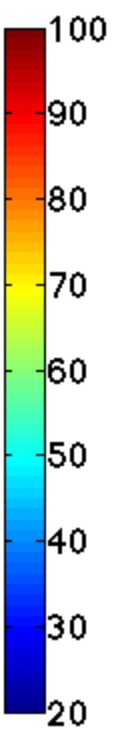} \vspace{3pt} \\
\includegraphics[width=5cm,height=5cm]{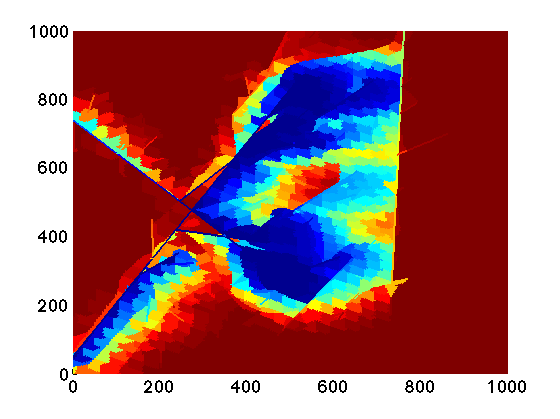}
\includegraphics[width=5cm,height=5cm]{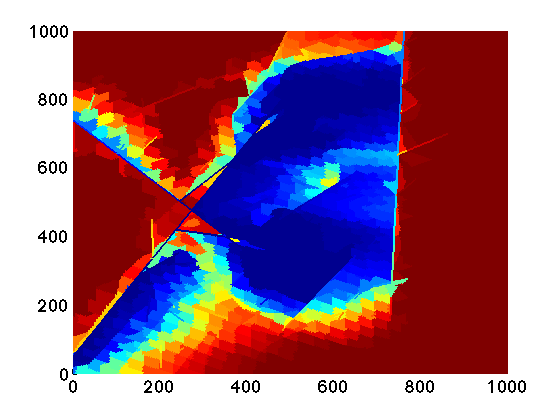}
\includegraphics[width=5cm,height=5cm]{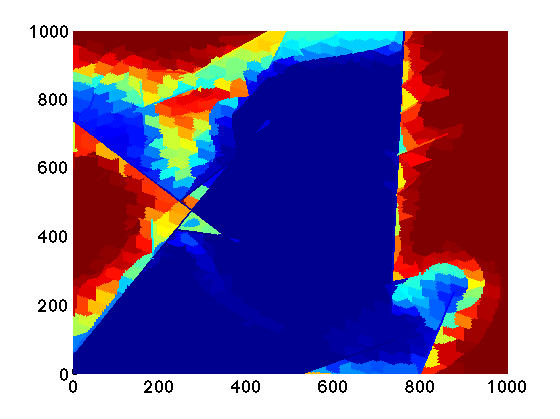}
\includegraphics[width=0.8cm,height=4.8cm]{Colorbar_20to100.png}
\includegraphics[width=5cm,height=5cm]{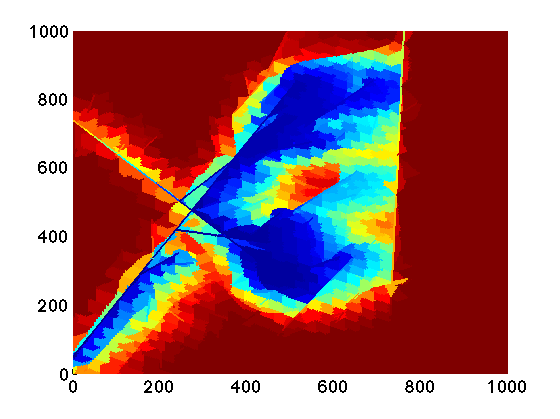}
\includegraphics[width=5cm,height=5cm]{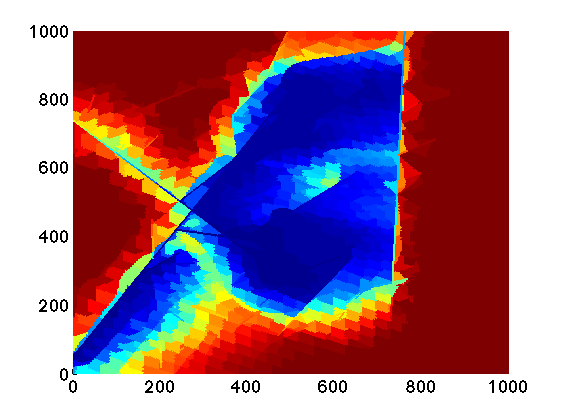}
\includegraphics[width=5cm,height=5cm]{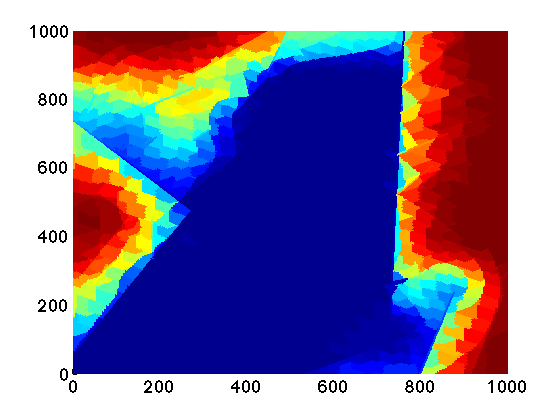}
\includegraphics[width=0.8cm,height=4.8cm]{Colorbar_20to100.png}
\caption{Time sequence of temperature profiles for reference solution (top) and upscaled solution with 1852 grid cells with smoothed basis (middle) and constant basis (bottom). The temperature profiles are plotted after $T = 0.5, 1,$ and  $5$ years, respectively, from left to right. Up to 111 iterations are used to smoothen the basis functions. The total relative energy errors at $T = 5$ years are $2.31 \cdot 10^{-2}$ for the upscaled method with smoothed basis and $3.68 \cdot 10^{-2}$ for the upscaled method with constant basis, both compared to the fine scale solution. The upscaling factor is 56.}
\label{time_sequence_ex3}
\end{figure*}

\begin{figure}
\center
\includegraphics[width=7cm,height=7cm]{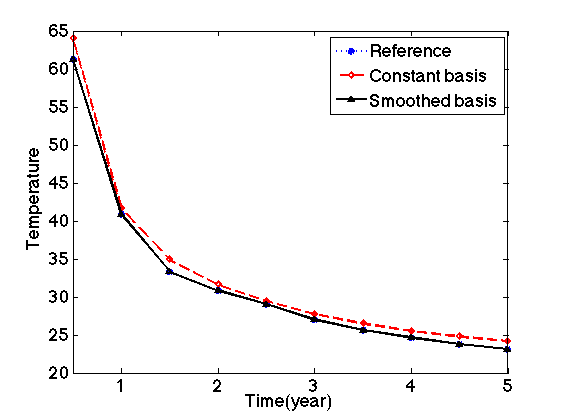}
\caption{Production temperature as a function of time for the simulations in section \ref{Pe_exp}. Temperature profiles are shown in figure \ref{time_sequence_ex3}.}
\label{1852_gc}
\end{figure}

In the simulations 1 $dm^2/s$ of water with temperature 20\degree C is injected into
the domain through a fracture near the bottom left corner and leaves through a
fracture near the upper right corner. Exact locations for injection and production
are shown in figure \ref{Fractures_2}. We use two computational grids, one fine
reference grid with $103893$ grid cells and an upscaled grid with $1852$ grid
cells, i.e the coarsening factor is $CF = 56$. The grid partitioning for the
coarse grid is shown in figure \ref{grid_partition}, to illustrate the
nonuniform grid structure. A time sequence of temperature profiles for
simulations using the fine grid and the upscaled grid with both smoothed and
constant basis functions are shown in figure \ref{time_sequence_ex3} at times $T
= 0.5, 1$ and $5$ years, respectively. 
For shorter times, advection is
dominating and both upscaled simulations capture fine scale behavior well, with
some additional numerical diffusion due to the coarse spatial discretization.
For longer times the difference between the smoothed basis and the constant
basis is more visible. The smoothed basis performs better compared to the
constant basis with respect to capturing heat conduction in the matrix, leading
to a less diffused temperature profile. 
As expected, the differences are largest towards the domain boundary, where the
flow rates are small, and conduction is the dominant transport process.
The production temperatures for both the
fine scale simulation and the upscaled simulations are shown in figure
\ref{1852_gc}. Both upscaled methods compare well to the fine scale reference
simulation, and with smoothed basis functions the upscaled production
temperature is closer to that of the reference solution compared to using the
constant basis.


\subsection{A Highly Complex Fracture Network}
\label{complex_network}

\begin{figure*}
\centering
\begin{overpic}[width=7cm,height=7cm]{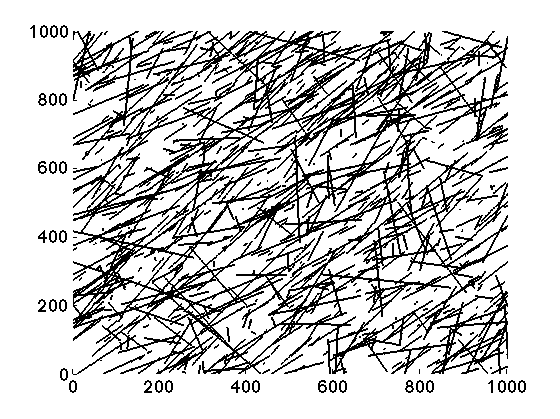} 
\linethickness{0.3mm}
\put(14,3){\vector(0,1){8}}
\put(10,0){\text{Injection}}
\put(99,91){\vector(-1,0){8}}
\put(78,94){\text{Production}}
\end{overpic}
\includegraphics[width=7.8cm,height=7cm]{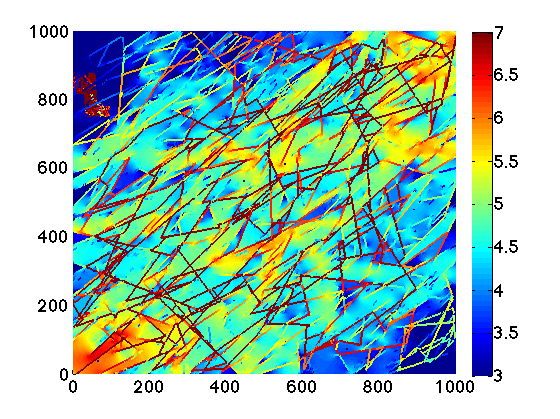} 
\caption{Fractures in the discretization for the simulations in section \ref{Pe_exp} (left) and the field with heat P\'eclet number per grid cell (right). The heat P\'eclet number plot is in logarithmic scale with base $10$. Injection and production locations are indicated by arrows in the fracture network plot.}
\label{Fractures_3}
\end{figure*}

As a final demonstration of the performance of the coarse discretization framework
we apply the method to a very challenging fracture network, shown in figure
\ref{Fractures_3}. The fractures are generated with Frac3D
\cite{SilberhornHemminger2002}, as one set of fractures with high fracture
density and the aperture is set to $a= 0.01$ m. All fractures are represented
both in the pressure grid and in the transport grid to demonstrate the
applicability of the method. For efficiency, disconnected fractures can be
excluded in the explicit representation and upscaled into the porous medium. We
consider a constant background permeability of $10^4$ mD approximating the
upscaled fractured media. The fracture permeability is modeled by equation
\eqref{frac_perm}, leading to a contrast between fracture permeability and
background permeability of almost six orders of magnitude. In figure
\ref{Fractures_3} the average heat P\'eclet numbers per fine grid cell are shown in
addition to the fracture network, as well as production and injection locations.
We observe that due to the high fracture density, there are only small matrix
regions with low flow rates.
The fine grid contains $355104$ grid cells and the upscaled grid $25616$ grid
cells, i.e. the coarsening factor is $CF=14$. The smoothening of basis functions is terminated after $45$ iterations. The injection rate is 5 $dm^2/s$ and the
injection temperature is 20\degree C. 

\begin{figure*}
\includegraphics[width=5cm,height=5cm]{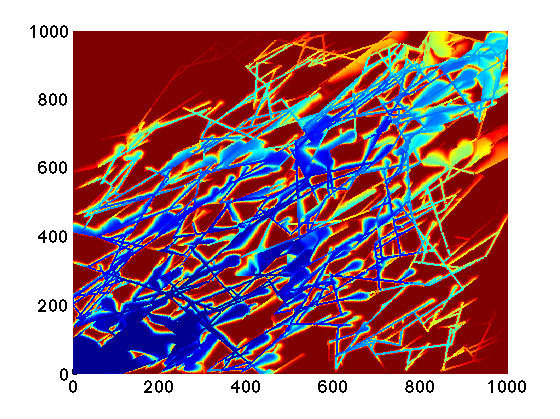} 
\includegraphics[width=5cm,height=5cm]
{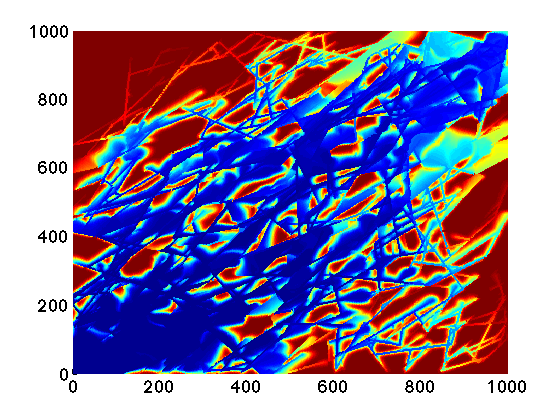} 
\includegraphics[width=5cm,height=5cm]{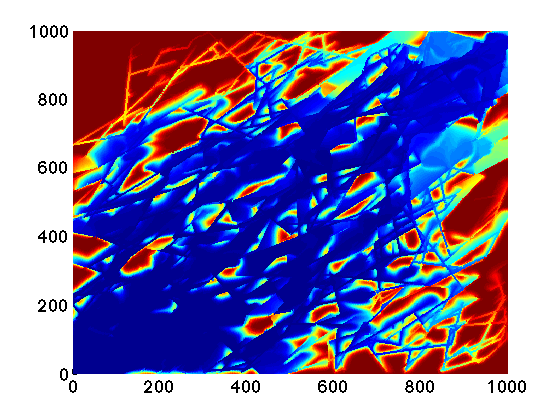} 
\includegraphics[width=0.8cm,height=4.8cm]{Colorbar_20to100.png}
\includegraphics[width=5cm,height=5cm]{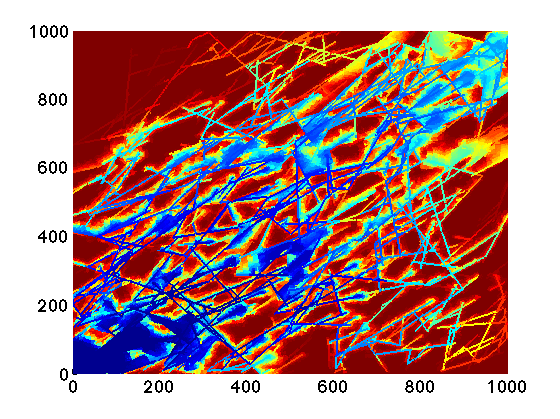} \hspace{6pt}
\includegraphics[width=5cm,height=5cm]{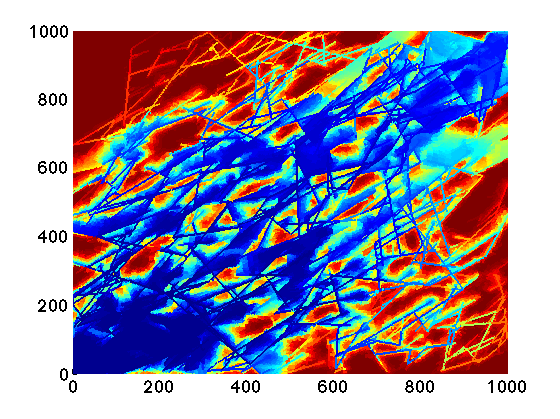} \hspace{6pt}
\includegraphics[width=5cm,height=5cm]{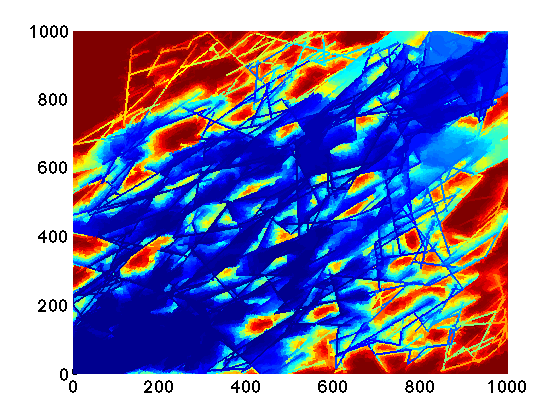} \hspace{6pt}
\includegraphics[width=0.8cm,height=4.8cm]{Colorbar_20to100.png}
\includegraphics[width=5cm,height=5cm]{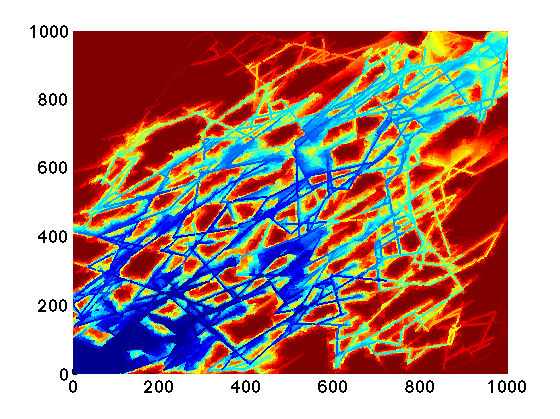} \hspace{8pt}
\includegraphics[width=5cm,height=5cm]{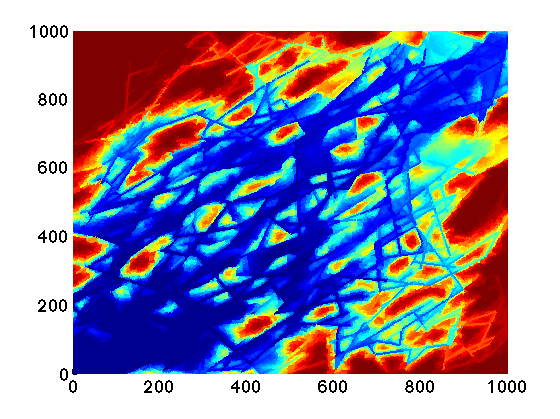} \hspace{10pt}
\includegraphics[width=5cm,height=5cm]{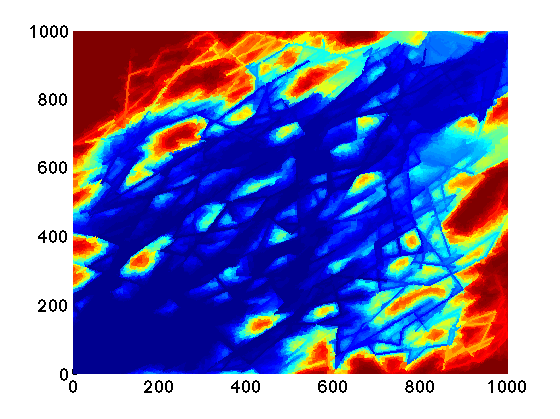}   \hspace{9pt}
\includegraphics[width=0.8cm,height=4.8cm]{Colorbar_20to100.png}
\caption{Fine scale and upscaled temperature profiles after $T=0.2, 0.6$, and $1$ years, respectively, shown from left to right. The fine scale simulation uses 355104 grid cells and the upscaled simulations uses a grid with with 25616 grid cells, i.e. the upscaling factor is $14$. The fine scale temperatures are shown in the top figures, temperatures with smoothed basis functions in the middle, and temperatures using constant basis functions in the bottom figures, respectively.  The fracture network used in the simulations is shown in figure \ref{Fractures_3}.} 
\label{complex_case_1_year}
\end{figure*}

\begin{figure}
\centering
\includegraphics[width=7cm,height=7cm]{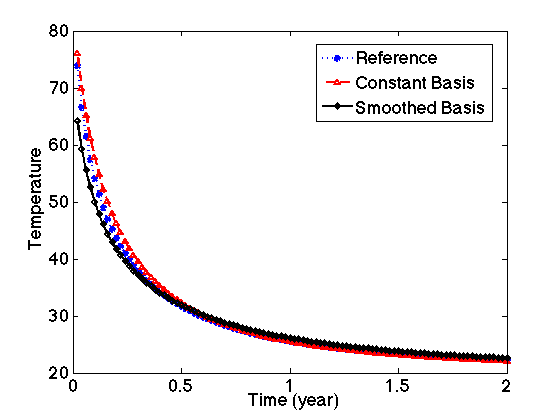} 
\caption{Production temperature as a function of time for the simulations in section \ref{complex_network}. Temperature profiles are shown in figure \ref{complex_case_1_year}.}
\label{temp_prod_well}
\end{figure}

\begin{figure}
\centering
\includegraphics[width=7cm,height=7cm]{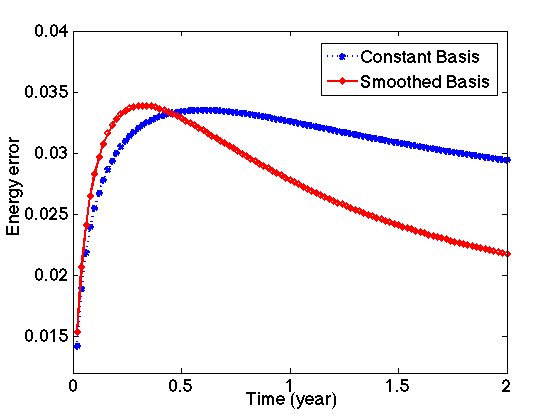} 
\caption{Reletive $\ell_2$ energy error over space as a function of time for upscaled simulations in section \ref{complex_network}, for a constant basis and a smoothed basis, respectively.}
\label{energy_error_over_time}
\end{figure}

Figure \ref{complex_case_1_year} shows the fine scale and upscaled temperature
profiles at $T = 0.2, 0.6$ and $1$ years, respectively. Both methods preserve
the fracture heterogeneity very well. Over longer times, it is clear from figure
\ref{complex_case_1_year} that the smoothed basis adds less diffusion compared
to the constant basis. Figure \ref{temp_prod_well} shows the production
temperature profiles, and we see that even for this highly complex fracture
network, the upscaled methods perform very well, with a similar asymptotic
behavior for both methods. This is also reflected by how the $\ell_2$ energy
errors over space changes over time, as shown in figure
\ref{energy_error_over_time}. As time progresses, the difference between the
$\ell_2$ energy errors for a smoothed basis and a constant basis increase in
favor for the smoothed basis.
Nevertheless, the numerical experiments clearly show that both methods can handle complex
fracture networks.

%



\section{Concluding Remarks}
\label{concl}
In this paper, we have presented an upscaling methodology for heat transfer in
geothermal fractured reservoirs, modeled by an advection-conduction equation for
the temperature. The methodology uses flow- and structure-based upgridding strategies well
suited for advection-driven processes. The upgridding is combined with a discretization of the coarse scale
conductive term based on an algebraic construction of basis functions. 

We present two discretizations of the coarse conductive term:
One based on piecewise constant functions, and a second formed by 
a the carefully constructed smoothing procedure of the constant functions.
While the former has a simple construction, the latter leads to a superior represenatation of heat transfer
via conduction from matrix to fractures. We demonstrate how to terminate the
smoothing iterations to avoid oscillations in the numerical solution.

Numerical simulations show that our upscaled framework produces accurate results compared
to fine scale numerical solutions for relatively large upscaling ratios,
i.e. at a significantly lower cost compared to fine scale simulations.
We also demontrate the applicability of the upscaling framework to a highly complex fracture network, and show that the structural heterogeneity of the fractured reservoir can be preserved.

Both the upgridding and the construction of basis functions are based on
algebraic approaches, and in principle extensions to three dimensions should
therefore be feasible.
However, the upgridding procedure will likely produce no less complex shapes of
the coarse cells, and the construction of smoothed basis functions may therefore
require further investigation to be robust in 3D.
On the other hand, constant basis functions should be readily applicable, and as
the numerical examples show will also give satisfactory accuracy in many cases.

In total, we believe that the present framework has the potential to
significantly facilitate field-scale simulation of heat recovery from fractured rocks.


%

\begin{acknowledgements}
This work is supported by VISTA project no. 6357.
\end{acknowledgements}



\end{document}